\documentclass{article}
\usepackage{graphicx} 
\usepackage{amsmath}
\usepackage{geometry}
\title{Bayesian competing risks survival modeling for assessing the
cause of death of patients with heart failure}
\author{Jesús Gutiérrez-Botella, Carmen Armero, Thomas Kneib, \\María Pata \& Javier García-Seara }
\date{24 Sept, 2024}

\begin{document}

\maketitle

\section*{Abstract}

Competing risks models are survival models with several  events of interest acting in competition  and whose occurrence is only observed for the event that occurs first in time. This paper presents a Bayesian approach to these models in which the issue of model selection is treated in a special way by proposing generalizations of some of the Bayesian procedures used in univariate survival analysis. This research is motivated by a study  on the survival of patients with heart failure undergoing cardiac resynchronization therapy,   a procedure which involves the implant of a device to stabilize the heartbeat. Two different causes of death have been considered: cardiovascular and non-cardiovascular, and a set of baseline covariates are examined in order to   better understand their relationship with both causes of death. Model  selection procedures and model checking analyses have been implemented and assessed. The posterior distribution of some relevant outputs   such as the overall  survival function, cumulative incidence functions, and transition probabilities have been computed and discussed.

\section{Introduction}

Cardiac resynchronization therapy (CRT) is an electrical intervention for heart failure patients. This treatment significantly enhances their quality of life, functional capacity, and reduces heart failure-related hospitalizations and cardiovascular mortality. It is  particularly indicated in patients with a left ventricular ejection fraction (LVEF)   less than or equal to 35\% and complex QRS width higher than 130 milliseconds despite optimized pharmacological therapy (Bristow \textit{et al.}, 2004; Cleland  \textit{et al.}, 2005; Moss \textit{et al.}, 2009).

LVEF  and left ventricular end-diastolic and end-systolic volumes (LVEDV and LVESV, respectively) are commonly used as clinical markers reflecting global left ventricular  systolic performance or   remodeling. CRT fosters reverse remodeling of the left ventricle, leading to decreased LVESV and improved LVEF. Clinical studies have established that an increase above  5\% in LVEF or a reduction of 15\% in LVESV are associated with enhanced survival (Cleland \textit{et al.}, 2013; Ghio \textit{et al.}, 2009). These benchmarks are commonly used to define a favorable response to CRT. Nevertheless, despite advances in technology and patient selection, around 30\% of patients are classified as non-responders to this therapy, a definition that might now be under debate owing to the observed benefits in lesser degrees of left ventricular remodeling compared to the natural course of the disease.

Different studies agree on reporting a mortality risk of approximately   10\% among patients receiving CRT, with heart failure being the leading cause (Marijon \textit{et al.}, 2015; Alvarez-Alvarez \textit{et al.}, 2021). In addition, some researchers have   identified male gender, advanced age, ischemic etiology, and severity of mitral regurgitation as indicators of mortality in this cohort (Cleland \textit{et al.}, 2005). However, the risk of mortality and heart failure-related hospitalizations fluctuates during follow-up. Assessing left ventricular (LV) reverse remodeling through longitudinal measurements of LV structure and function over time in heart failure patients presents a more comprehensive method for stratifying cardiovascular mortality and heart failure decompensation risk (Alvarez-Alvarez \textit{et al.}, 2021).

The objective of this paper is to investigate baseline predictors of cardiovascular and non-cardiovascular mortality as well as to assess the survival times in heart failure patients undergoing CRT, knowing that their death may eventually be due to cardiovascular or non-cardiovascular causes.

The statistical framework for the aforementioned objectives is survival analysis. In particular, competing risks models in which the event of interest, in our case death, can occur from a variety of competing causes. The statistical literature is abundant regarding procedures for modeling competing survival scenarios (Putter \textit{et al.}, 2007; Chen \textit{et al.}, 2014). In fact, competing risks models have been widely applied to numerous settings, such treatment evaluation in clinical trials (Ameis \textit{et al.}, 2024), breast cancer (Basu and Tiwari, 2009) or hepatic diseases (Yu and Huang, 2023), among other relevant clinical topics.

In our case, we will focus on cause-specific risk modelling to quantitatively assess the behaviour of the survival process through the cause-specific hazard function for each cause of death, the overall survival function for the time in which death occurs, and the cumulative incidence function of that time due to each of the causes considered  among others  (Chen \textit{et al.}, 2014). We also incorporate  a look at the modelling from the point of view of multistate models (Andersen \textit{et al.}, 2001). These are stochastic processes of continuous time and discrete state space representing the different health conditions considered. We discuss the transition probabilities between the different states and specially conditional transition probabilities when it is known that a patient remains alive a certain time after the CRT treatment.  We will work within the Bayesian statistical methodology. This  allows quantifying all sources of uncertainty associated with a statistical problem in a natural way through probability distributions. 

The rest of the paper is organized as follows. Section 2 presents a retrospective follow-up cardiac resynchronization therapy study and a brief description of the subsequent dataset. Section 3 contains general information about the competing risks model, the main elements of the  cause-specific risk  approach, and the general Bayesian inferential process.  Subsection 3.3, devoted to model uncertainty, presents a special variable selection scenario. In this topic, we  extend the proposal by García   García-Donato \textit{et al} (2023) in Cox regression models for computing Bayes factors  to competing risks models and  employ spike-and-slab selection methods. In addition, we  also expand the definition of conditional predictive ordinate (CPO)  (Ibrahim \textit{et al.}, 2001) to the general competing risks framework which is needed in order to apply the logarithm of the pseudomarginal likelihood as a tool for model assessment.  Section 4   returns to the study of CRT therapy and applies the procedures and proposals of Section 3 to the data of this problem. Finally, Section 5  includes some conclusions.

\section{The cardiac resynchronization therapy study}

This is a  retrospective follow-up study including 296  patients undergoing CRT,  defibrillator  or pacemaker, under standard clinical indications in a single tertiary cardiac institution in Galicia (Spain) between  January 2005 and April 2015. All patients demonstrated heart failure symptoms with ischaemic or non-ischaemic cardiomyopathy, decreased LVEF ($\leq$35$\%$) and prolonged QRS duration ($\geq$120 ms) at the time of the implantation. The New York Heart Association (NYHA) functional classification was used to assign them to three groups:   class II, when presenting symptoms of heart failure with moderate exertion; class III, with minimal exertion; and class IV  at rest. This study was conducted in accordance with the Principles of the Declaration of Helsinki and was approved in 2017 (code: 2017/171) by the Clinical Research Ethics Committee (CEIC) of Galicia.

Clinical follow-up was completed in all patients, from the time each individual  underwent CRT implantation (time zero) until death, cardiovascular (CV) or non-cardiovascular (N-CV), or until the date of the end of the follow-up.  Survival times were observed in the first case and were right-censored in the latter.   The median and interquartile ranges of survival times were 2.75 and (1.2, 5.2) years, respectively. Survival times and times-to-event for each cause of death densities are represented in Figure 1.

\begin{figure}[htb]
\begin{center}
\includegraphics[scale=0.6]{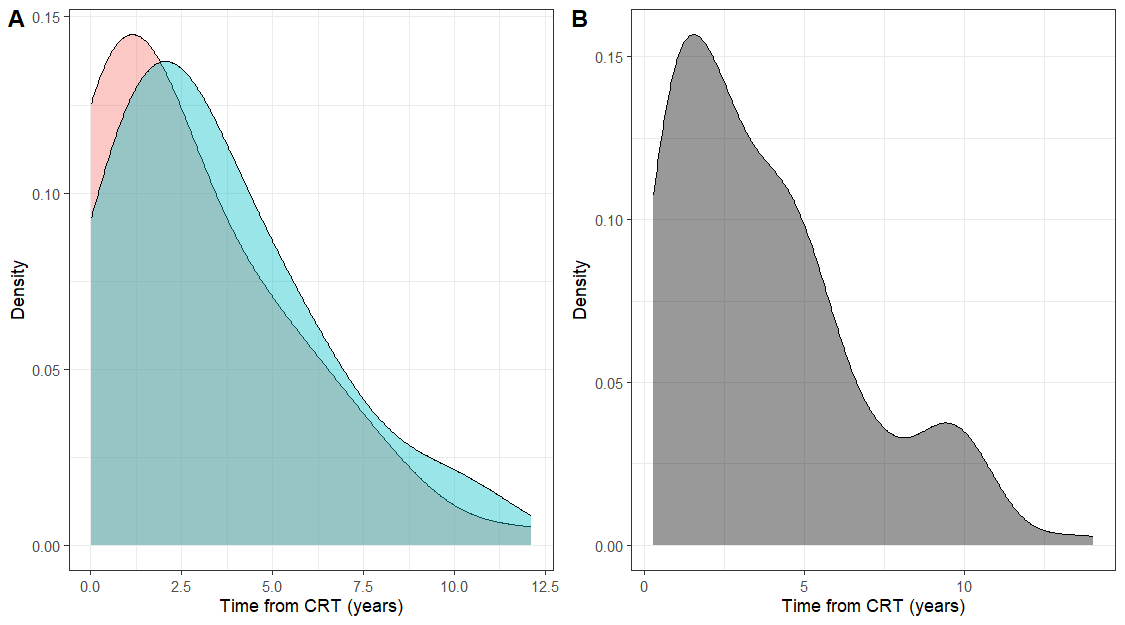}
\caption{Distributions of times-to-event. In (A) times to cardiovascular and non-cardiovascular deaths in red and blue, respectively; in (B) times to censoring.}
\end{center}
\label{figure:0}
\end{figure}

Concrete cause of death was registered for each patient. CV death involves causes as stroke and heart ischemic attacks; while NCV deaths include different causes such as bleedings, cancer or infections. In all cases, the exact cause of death was effectively registered and grouped into CV or NCV.

Patients were individually followed up  in the hospital's arrhythmia unit to verify the proper functioning of the device. Subsequently, visits were conducted to check the battery status, the patient's rhythm, stimulation percentage, stimulation and sensing thresholds in each chamber, and electrode impedances. Arrhythmic episodes and corresponding electrograms were reviewed, with different adjustments made based on each patient's condition. Long-term follow-up included a review of the medical history, which encompassed the development of atrial fibrillation, cardiovascular admissions - including hospitalizations due to heart failure - and mortality, specifying the cause of death. Additionally, the duration of the subsequent QRS on surface   electrocardiogram  and the evolution of echocardiographic variables were collected through repeated measurements throughout the follow-up period.

A retrospective review was also conducted for each of the patients in the study, including electronic medical reports as well as older records available only in paper format. All medical interventions and hospital records were documented. Vital status was ascertained through telephone calls in the absence of medical reports during follow-up. Furthermore, pharmacological treatment was assessed using the IANUS electronic registry (a database containing patient medical history, including reviews and treatments).


The covariates considered in the study were  selected by the team cardiologist. They  included age and sex of all patients at the time of implantation, etiology of heart failure (ischemic vs. non-ischemic or dilated), NYHA functional class, creatinine level  in blood as a biomarker of renal function, QRS width at the moment of implantation, the type of device which has been implanted (pacemaker or implantable cardioverter-defibrillator ) and the possible occurrence of a  left bundle branch block (LBBB) that generates disruptions to the electronic impulse that causes the heart to beat.   Additionally, the history of previous atrial fibrillation (Previous AF)  has been registered.



\begin{table}[htb]
\begin{center}
\caption{Descriptive summary of study covariates for censored individuals, for those who died, from CV or N-CV causes, and for all of them together. In the case of quantitative covariates, the sample mean and standard deviation are presented; for categorical variables, the number of patients in the sample in each category and their percentage are reported.  }
\begin{tabular}{lllll}
\hline
 & \textbf{Censored} & \textbf{CV Death} & \textbf{N-CV Death} & \textbf{Total}\\
\hline
\vspace{-0.3cm} \\
$n$ & 203 & 55 & 38 & 296\\
\vspace{-0.2cm} \\
\multicolumn{5}{l}{\textbf{Categorical covariates.}}\\
\vspace{-0.4cm} \\
CRT Device&  &  &  & \\
- Pacemaker & 87 (42.86\%) & 36 (65.45\%) & 17 (44.74\%) & 140 (47.30\%)\\
- Defibrillator & 116 (57.14\%) & 19 (34.55\%) & 21 (55.26\%) & 156 (52.70\%)\\
\vspace{-0.3cm} \\
Etiology &  &  &  & \\
- Ischemic & 62 (30.54\%) & 28 (50.91\%) & 16 (42.11\%) & 106 (35.81\%)\\
- Dilated & 141 (69.46\%) & 27 (49.09\%) & 22 (58.89\%) & 190 (64.19\%)\\
\vspace{-0.3cm} \\
LBBB & & & &  \\ 
- Yes & 80 (39.41\%) & 26 (47.27\%) & 12 (31.58\%) & 118 (39.86\%) \\
- No & 123 (60.59\%) & 29 (52.73\%) & 26 (68.42\%) & 178 (60.14\%) \\
\vspace{-0.3cm} \\
NYHA &  &  &  & \\
- Class II & 70 (34.48\%) & 3 (5.45\%) & 4 (10.53\%) & 77 (26.01\%)\\
- Class III & 126 (62.07\%) & 43 (78.18\%) & 33 (86.84\%) & 202 (68.24\%)\\
- Class IV & 7 (3.45\%) & 9 (16.36\%) & 1 (2.63\%) & 17 (5.74\%)\\
\vspace{-0.3cm} \\
Previous AF & & & &  \\
- Yes & 83 (40.89\%) & 31 (56.36\%) & 24 (63.16\%) & 138 (46.62\%) \\
- No & 120 (59.11\%) & 24 (43.64\%) & 14 (36.84\%) & 158 (53.38\%)\\
\vspace{-0.3cm} \\
Sex &  &  &  & \\
- Man & 150 (73.89\%) & 48 (87.27\%) & 29 (76.32\%) & 227 (76.69\%)\\
- Woman & 53 (26.11\%) & 7 (12.73\%) & 9 (23.68\%) & 69 (23.31\%)\\
\vspace{-0.2cm} \\
\multicolumn{5}{l}{\textbf{Quantitative covariates.}}\\
\vspace{-0.4cm} \\
Age & 68.58 (10.00) & 72.27 (7.73) & 74.18 (7.15)  & 69.99 (9.51) \\
\vspace{-0.3cm} \\
Creatinine & 1.24 (0.45) & 1.70 (0.84) & 1.39 (0.49)  & 1.34 (0.57) \\
\vspace{-0.3cm} \\
QRS & 161.22 (24.88) & 162.51 (31.71) & 167.89 (25.01)  & 162.32 (26.26) \\
\hline
\label{tab:1}
\end{tabular}
\end{center}
\end{table}

Table \ref{tab:1} shows a basic description of the main variables in the study for censored individuals  and for the people who died, both from CV and N-CV causes. It also includes a fourth column accounting for all the data. In the case of the categorical variables, sex, device, etiology, LBBB, NYHA and previous atrial fibrillation,   the frequency and percentage associated with each category of the variable is shown. The majority (203) of the 296 individuals  were censored for the event death. Of the remaining 93 persons who died, 55 died from CV causes and 38 from N-CV causes. 76.69 $\%$ of the patients were men. This proportion remains the same in the group of censored individuals and in the group of individuals who died of N-CV causes, but increases to around 87 $\%$ in the people who died of CV causes. For the continuous covariates, age, creatinine, and QRS, the mean and standard deviation are shown.  The mean and standard deviation of creatinine in the group of people who died due to CV problems is higher than that of the other two groups, as well as, although slightly, the presence of LBBB and QRS. On the other hand, in the censored group there is a much lower proportion ($40.89$\%$)$ of previous AF than in the CV death group (56.36$\%$) and N-CV death (63.16$\%$). The proportion of people who have had a pacemaker implanted is much higher in the group of people who died due to CV causes. The remaining two groups maintain similar values, 42.86$\%$ and 44.74$\%$ in the group of those censored and those who died from N-CV causes. The etiology of the disease, ischemic and dilated, is also distributed approximately equally in the group of those who died of CV causes, but in the group of those censured it was about 30$\%$ and 70$\%$ and in the group of those who died of N-CV causes 42$\%$ and 58$\%$, respectively. 

The standard Kaplan-Meier procedures for describing survival probabilities do not provide proper results in the context of competing risks (Pepe and Mori, 1993).  In these settings, an alternative measure is the cumulative incidence function for each cause, defined as the cumulative probability   of failure from a specific cause before a certain time. Figure 2 presents the cumulative incidence function
for CV and N-CV death computed from the R package \texttt{cmprsk} (Gray, 2022). Specifically, this cumulative incidence function was calculated as described in Gray, 1998. CV and N-CV death behave quite similarly, although cardiovascular death is always associated with a higher probability. It is important to note that after five years of the CRT intervention the difference between the two probabilities increases: between 0.15 and 0.20 for    N-CV death and between 0.20-0.25 for CV death. From 7-8 years onwards there is an abrupt rise in CV death, while CV remains at around 0.2. The small number of patients we have from 10 years onwards does not produce consistent informative signals.

\begin{figure}[htb]
\begin{center}
\includegraphics[scale=0.4]{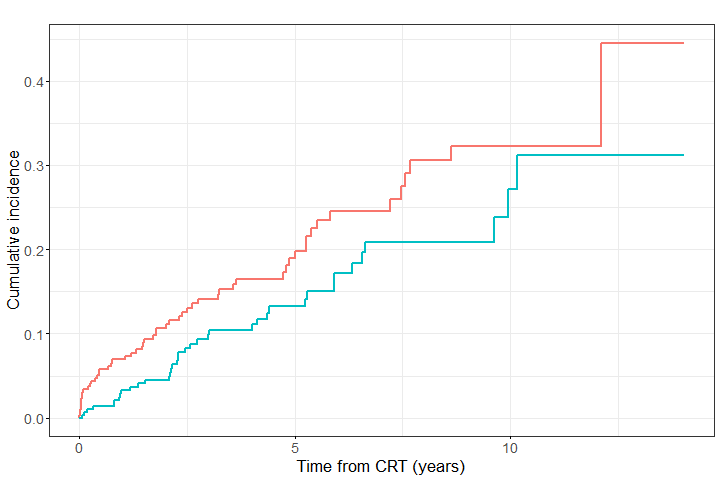}
\caption{Estimation of the cumulative incidence  function for time to CV (red) and non-CV (blue) death. }
\end{center}
\label{figure:1}
\end{figure}

\section{Bayesian competing risks modelling}

\subsection{Competing risks sampling model}

Competing risks models are survival models in which the initiating event of the process contemplate different terminating events, which compete with each other to be the event that eventually occurs. 
There are different models suitable for analyzing competing risks problems, being the most used the multivariate time-to-events model (Gail, 1975), the sub-distribution model (Gray, 1988) or the cause-specific hazards model (Gaynor \textit{et al.}, 1993). The most popular, and the one we will use in our study, is the latter: the cause-specific hazards modelling. An updated review on this models and some applications can be found in Chen \textit{et al.}, 2014.

We assume $K$ terminating events or causes of failure. We define  $T_k$ as the random variable that describes the time to the terminating event $k$,   and $\delta$ as the indicator variable for the event that has occurred. In particular,     $\delta=k$ when the event $k$ has occurred and $\delta=0$ indicating the situation in which no event of interest has been recorded during the study period and, therefore, we have   right censored information. Let $T= \mbox{min}\{T_1, \ldots, T_K\}$. Cause-specific hazard modelling    uses as a basic concept the joint distribution of $T$ and $\delta$ through the so-called  cause-specific hazard function, which represents the hazard of failing from a given event  in the company of all other competing events. The cause-specific hazard function for $T$ due to cause $k$ is defined as
  \begin{equation}
h_{k}(t \mid \boldsymbol\theta_{k})=  \lim_{\varDelta t \to  0} \frac{P(t \leq T < t + \varDelta t, \delta=k \mid T \geq t,\, \boldsymbol \theta_k)}{\varDelta t},
\label{eqn:specific}
\end{equation}
where $\boldsymbol \theta_k$ is the subsequent parametric vector. It is important to remember that our inferential framework is Bayesian statistics. Consequently, the uncertainty associated with the parameters will be expressed in probabilistic terms and for this reason we will express in terms of conditional probability distributions all probabilistic elements, such as $h_k(t \mid \boldsymbol{ \theta_k})$,   that depend on  parameters. 

 The most common modeling of the cause-specific hazard function for cause $k$ is through a   proportional hazards  model  as follows
\begin{equation}
h_{k}(t\mid \boldsymbol \theta_{k})=h_{0,k}(t\mid \boldsymbol \phi_{k})\:\exp\{\boldsymbol x^{\top}_k \, \boldsymbol \beta_{k} \}, 
\end{equation}
where $h_{0,k}(t\mid \boldsymbol {\phi_{k}})$ is a baseline hazard function that depends on time and a parametric vector $\boldsymbol \phi_k$. It expresses the shape of the cause-specific hazard function. The vector  $\boldsymbol \beta_k$ represents the regression coefficients associated to covariates $\boldsymbol x_k$ of covariates, and $\boldsymbol \theta_k=(\boldsymbol \phi_k, \boldsymbol \beta_k)$. Baseline hazard functions can be established through parametric or semiparametric approaches. Parametric models generate functions with rigid shapes that cannot admit irregular behavior. On the other hand, semi-parametric modeling provides very flexible functions but suffer from   overfitting and instability (Ibrahim \textit{et al.}, 2014).

Outputs of interest in competing risks models  include the overall survival function and the cumulative incidence function from each competing event $k$. The overall survival function for $T$ is 
\begin{equation}
S(t \mid \boldsymbol \theta) = P(T>t \mid \boldsymbol \theta)= \exp\left\{ -\sum_{k=1}^{K}\int_{0}^{t}h_{k}(u \mid \boldsymbol \theta_k)\:du\right\}, 
\label{eqn:overall}
\end{equation}
where now $\boldsymbol \theta=(\boldsymbol \theta_1, \ldots, \boldsymbol \theta_K)$. 

The cumulative incidence function of $T$ due to  cause $k$  describes the time-dependent probability that an individual experiences event $k$ in the presence of all the competing events as a function of time:
\begin{equation*}
F_{k}(t \mid \boldsymbol \theta )=P(T\leq t,\:\delta=k \mid \boldsymbol \theta)=\int_{0}^{t}h_{k}(u \mid \boldsymbol {\theta_k})\:S(u \mid \boldsymbol \theta)\:\mbox{d}u.
\end{equation*}

This function provides the probability associated with the occurrence of each of the competing events, $P(\delta =k \mid \boldsymbol \theta)=  \lim_{ t \to  \infty} F_k(t \mid \boldsymbol \theta)$.

Competing risks models can also be seen as a particular multi-state model. In general, a multi-state process is a continuous time stochastic process 
$\{Z(t), t \geq 0\}$ with a finite   state space $\Omega=\{0,1,\ldots, K\}$, where $Z(t)=k$  indicates that the process is in state $k$ at time    $t$, or in terms of  survival analysis that event $k$ has occurred at some time instant less than $t$.  The initial distribution of the process is $P(Z(0)=0)=1$ and relevant transition probabilities between states are
\begin{align*}
p_{(0,0)}(0,t\mid \boldsymbol{\theta}) & =P(Z(t)=0 \mid Z(0)=0, \boldsymbol \theta)=P(T > t \mid \boldsymbol \theta)= S(t \mid \boldsymbol \theta), \\
p_{(0,k)}(0,t\mid \boldsymbol \theta) & =P(Z(t)=k \mid Z(0)=0, \boldsymbol \theta)=P(T \leq t, \delta= k \mid \boldsymbol \theta)=F_{k}(t \mid \boldsymbol \theta ), \,\,k\neq 0.
\end{align*} 
In many instances, for example in disease monitoring, it is often important for both the physician and the patient to assess the risks associated with disease progression not only from the beginning of the monitoring but also from subsequent periods. In these cases, the relevant transition probabilities are: 

\begin{align*}p_{(0,0)}(s,t\mid \boldsymbol \theta)  = &\,
\frac{P(Z(t)=0, Z(s)=0 \mid \boldsymbol \theta) }{P(Z(s)=0 \mid \boldsymbol \theta)} =           \frac{S(t \mid \boldsymbol \theta)}{S(s \mid \boldsymbol \theta)},\,t \geq s,\\
p_{(0,k)}(s,t\mid \theta)  = & \,
\frac{P(Z(t)=k, Z(s)=0 \mid \boldsymbol \theta) }{P(Z(s)=0\mid \boldsymbol \theta)} =           \frac{P(s<T<t, \delta=k \mid \boldsymbol \theta)}{S(s \mid \boldsymbol \theta)} \\ = & \, \frac{F_k(t \mid  \boldsymbol \theta)-F_k(s \mid \boldsymbol \theta)}{S(s \mid \boldsymbol \theta)},\,t \geq s, \,k\neq 0.  \end{align*}

\subsection{Prior and posterior distributions}

  Bayesian models include a sampling model and a prior distribution $\pi(\boldsymbol \theta)$ for the unknown parameters that describes how much, little or no information is available about $\boldsymbol \theta$. We will work in a general scenario of prior independence between the parameters associated with the different competing risks as well as between the parameters governing the time-dependent structure of each hazard function and the regression coefficients of the variables, $\pi(\boldsymbol \theta)=\prod_{k=1}^K\, \pi(\boldsymbol \phi_k, \boldsymbol \beta_k)= \prod_{k=1}^K\, \pi(\boldsymbol \phi_k) \,\pi(\boldsymbol \beta_k)$.

  The database is built on the basis that    survival times due to one of the causes provides censored information on the survival times of the other competing causes. The information provided by the data $\mathcal D$ is described in terms of the likelihood function of $\boldsymbol \theta$, $L(\boldsymbol \theta)$, which is calculated through the hazard function (Chen \textit{et al.}, 2014). The prior distribution and the likelihood function are combined to obtain the posterior distribution of $\boldsymbol \theta$, $\pi(\boldsymbol \theta \mid \mathcal D)$. This distribution is not analytical  but an approximate sample  $\{\boldsymbol \theta^{(1)}, \ldots, \boldsymbol \theta^{(M)}\}$ of such a distribution can be obtained through Markov chain Monte Carlo, MCMC, methods. This distribution includes all available information and is the starting point for the analysis of the problem. 
  
  The overall survival function, cumulative incidence function, probability associated with the occurrence of each competing event as well as transitions probabilities presented before are relevant outputs that depend on $\boldsymbol \theta$. Consequently, we can use the posterior approximate sample from  $\pi(\boldsymbol \theta \mid \mathcal D)$ to generate approximate random samples for their subsequent posteriors,  $\pi(S(t \mid \boldsymbol \theta) \mid \mathcal D)$, $\pi(F_k(t \mid \boldsymbol \theta) \mid \mathcal D)$, and $\pi(p_{(u,v)}(s,t \mid \boldsymbol \theta) \mid \mathcal D)$, respectively.
  
\subsection{Model uncertainty}

We assume uncertainty in the model, both in the specification of the baseline hazard function and in the set of covariates. In our problem, our main interest is in delving into the role of the covariates in the model and evaluating the real information that each of them provides about the survival times of interest.  Model selection is a challenging problem in statistics from a methodological point of view and of great importance in modeling real problems. In current regression studies, increasingly larger sets of covariates are available in which covariates that really have a relevant impact on the variable of interest are mixed with covariates that barely provide information about the problem or whose information, although important, is redundant. Identifying the really valuable variables and discarding those that are not is a really necessary process. Regression models with many covariates are complex and rigid models that generate estimates with a lot of uncertainty due to situations of multicollinearity. They    have a limited predictive ability.

Statistical literature is very rich in many and different proposals that address model uncertainty. We  extend the proposal of García-Donato \textit{et al} (2023) in Cox regression  to competing risks models, and employ spike-and-slab methods for variable selection.
Both procedures use a strategy whose first stage proposes a variable selection method and a subsequent analysis of the estimated model.

The approach proposed in García-Donato \textit{et al.} (2023)  is based on computing the posterior probability associated to each of the possible models (model space), from the null model containing no covariates to the full model with all covariates. The calculation of such   posterior  distribution uses a discrete uniform prior distribution over the model space and a multivariate normal  distribution as the prior for  the regression coefficients
$\pi(\boldsymbol \beta_k )= \mbox{N}(\boldsymbol 0, \,\Sigma),$
where $\Sigma^{-1}$ is expressed in terms of the precision matrix of the covariates. This variance-covariance matrix   is  based on the literature of   conventional priors and the expected Fisher information matrix evaluated on the null model.      Model search in the different $2^{p_k}$ models, where $p_k$ is the number of covariates in the cause-specific hazard for cause $k$, $k=1, \ldots, K$, is based on Gibbs sampling algorithm and the Rao-Blackwellized estimator (Gelfand and Smith, 1990; Ghosh and Clyde, 2011)   is used for the estimation of the inclusion probability associated to each covariate.

The second selection method we will follow is based   on spike and slab prior distributions for the set of regression coefficients. In this approach, the  prior distribution associated to the 
$\boldsymbol \beta_k$ regression coefficients associated to cause $k$ is expressed as (Mitchell and Beauchamp, 1988)

\begin{align}
\pi(\boldsymbol \beta_{k}\mid \boldsymbol \gamma) & = \mbox{$\prod$}_{j=1}^{p_k} \,[(1-\gamma_j) I_0(\beta_{kj})+\gamma_j \pi(\beta_{kj})],  \nonumber \\
\pi(\boldsymbol \gamma \mid \eta) & = \mbox{$\prod$}_{j=1}^{p_k} \,\eta^{\gamma_{j}}(1-\eta)^{1-\gamma_j}, \nonumber \\
 \eta & \sim\mbox{Be}(\alpha_1, \alpha_2),
\label{eqn:spike}
 \end{align}
 where $ I_0(\beta_{kj})$ is a mass point at zero  that accompanies the spike, $\pi(\beta_{kj})$ is a wide density, and $\boldsymbol \gamma$ is a $\{0,1\}$ vector which defines the different  $2^{p_{k}}$ models.

 The two procedures we have used look at the covariates but are not designed, especially the first one, to assess the baseline hazard function. In this sense, as a complement to  model selection we focused   on   other criteria to assess the performance of the model as a whole,  with  covariates and a  baseline hazard function. The literature on this subject is extensive. We have selected  the Deviance Information Criterion (DIC) (Spiegelhalter \textit{et al.}, 2002) and the Watanabe-Akaike Information Criterion (WAIC)  (Watanabe, 2010). both based on information measures and the logarithm of the pseudomarginal likelihood (LPML) (Ibrahim \textit{et al.}, 2001) based on the predictive power of the model. 

DIC measures the fit of the data to the model via the deviance with a penalty for the complexity of the model, the effective number of parameters. This criterion has been a very popular procedure in the Bayesian world due to its practical relationship with MCMC procedures to approximate the posterior distribution and its good performance, basically in non-hierarchical scenarios as it is our case. Smaller values of DIC indicate a better-fitting model. DIC is defined as $\text{DIC}=D(\bar\theta)+2pD$, where $D(\theta)$ is the deviance of the posterior distribution of $\theta$, and $pD$ is the effective number of parameters and it is calculated as $pD=\text{E}[D(\theta)]-D(\bar\theta)$.

WAIC  criterion   is an extension of the Akaike Information Criterion  with a fully Bayesian conception.  Its definition is similar to the DIC, as it assesses the model fit and penalizes models with a high number of parameters. WAIC has been calculated as $\text{WAIC}=-2\sum_{i=1}^n \log p_{post(y_i)}+2pD$, where $\sum_{i=1}^n \log p_{post(y_i)}$ is the sum of predictive density for each data point and $pD$ the number of effective parameters. Although in the context of non-hierarchical modeling like this one DIC and WAIC should provide similar information, we have obtained both indexes for the sake of completeness. As expected, they have always reached similar conclusions.

LPML is based on the Conditional Predictive Ordinate  (CPO) associated to each observation  (Ibrahim \textit{et al.}, 2001). It is defined as
$$\mbox{LPML}=\sum_{i=1}^n\,\mbox{log}(\mbox{CPO}_i),$$
where CPO$_i$ for observation $y_i$ in general univariate   scenarios without censoring is defined in terms of the marginal posterior predictive distribution 

$$   f(t \mid \mathcal D^{-(i)})= \int\,f(t \mid \boldsymbol \theta)\,\pi(\boldsymbol \theta \mid \mathcal D^{-(i)} ) \,\mbox{d}\boldsymbol \theta, $$

\noindent where  the $\mathcal D^{(-i)}$ denotes the data without the $i$th observation (leave-one-out procedure), and  $f(t \mid \boldsymbol \theta)$ is the subsequent  probability sampling density. In the case of univariate survival times with right censoring, the CPO associated to observation $t_i$ is the value of that posterior marginal distribution evaluated at the observation $t_i$ if this time is observed and the posterior probability $P(T_i>t_i \mid \mathcal D^{(-i)})$ when $t_i$ is censored (Laud, 2014). Large CPOs values support the selected model because indicate a good tuning between the data and the model. 

The above definition of CPO does not apply to competitive risk models in which there are different event causes and censoring.  We propose an extension of the   definition of the CPO associated to the $i$-th observation $t_i$  in a competing risks model as follows: 
 \begin{itemize}
 \item Individual $i$ has not experienced  any of the events in the model during the period of the study. Consequently, observation $t_i$ for individual $i$ is censored for all competing events ($\delta_i=0$). Its   associated CPO is the value at $t_i$ of the posterior predictive distribution 
 $$  P(T > t \mid \mathcal D^{-(i)}) = S(t \mid \mathcal D^{-(i)})= \int S(t \mid \boldsymbol \theta)\, \pi(\boldsymbol \theta \mid \mathcal D^{-(i)})\, \mbox{d}\boldsymbol \theta,$$
 
\noindent where $S(t \mid \boldsymbol \theta)$ is the overall survival function in(\ref{eqn:overall}).  
 
\item Individual $i$ has experienced the event $k$ in time $t_i$ within the period of the study. In this case the CPO associated to this observation is the value at $t_i$ of the posterior predictive subdensity function 
$$\ f_k(t  \mid \mathcal D^{-(i)})=   \int f_k(t  \mid \boldsymbol \theta)\, \pi(\boldsymbol \theta \mid \mathcal D^{-(i)})\, \mbox{d}\boldsymbol \theta  =\int h_k(t  \mid \boldsymbol \theta)\, S(t  \mid \boldsymbol \theta)\,\pi(\boldsymbol \theta \mid \mathcal D^{-(i)})\, \mbox{d}\boldsymbol \theta,$$ 
 
\noindent where $f_k(t \mid \boldsymbol \theta)= h_k(t  \mid \boldsymbol \theta)\, S(t  \mid \boldsymbol \theta)$ is known as the subdensity  sampling function for time-to-event $k$, defined in terms of the  cause-specific hazard function  for cause $k$ in (\ref{eqn:specific}) and the overall survival function (\ref{eqn:overall}).

     \end{itemize}

CPOs and LPML can be used to compare the predictive performance of different models as well as to search for outliers.  We will use an approximation for the calculation of the CPOs based on the MCMC outputs  (Ntzoufras, 2008).

\section{Bayesian competing risks analysis in action: Cardiovascular and non-cardiovascular death }

\subsection{Model specification}

We return to the cardiac resynchronization therapy study, whose data we will analyse using a Bayesian competing risks model with  two competing events, CV and N-CV death. Each patient undergoing CRT (initial state) can access any of them after having  surgery. We define $T_C$ and $T_{NC}$  as the survival time associated with CV and N-CV death, respectively. We express the cause-specific hazard function for CV and N-CV death   in terms of a Cox proportionals hazards modelling,
\begin{align*}
    &h_{c}(t \mid \boldsymbol \phi_c,\, \boldsymbol \beta_c)= \, h_{0,c}(t \mid \boldsymbol \phi_c)\, \exp\{\boldsymbol x^{\top}_{c}\boldsymbol \beta_{c}\}, \nonumber \\
     &h_{nc}(t \mid \boldsymbol \phi_{nc},\, \boldsymbol \beta_{nc})= \, h_{0,nc}(t \mid \boldsymbol \phi_{nc})\, \exp\{\boldsymbol x^{\top}_{nc} \,\boldsymbol \beta_{nc}\},
\end{align*}
where $ h_{0,c}(t \mid \phi_c)$ ($h_{0,nc}(t \mid \phi_{nc})$) is the baseline hazard function for cardiovascular (non-cardiovascular) death and the element with covariates is :
\begin{align*}
 &\boldsymbol x^{\top}_{c}\boldsymbol \beta_{c}=  \,\,\beta_{c.1}I_{Man} +   \beta_{c.2}I_{Pace}   +\beta_{c.3}I_{Ische} +  \beta_{c.4}I_{III} +     \beta_{c.5}I_{IV}  \nonumber \\
& \hspace*{0.85cm} + \,\,\beta_{c.6}\,\mbox{Age}+  \beta_{c.7}\,\mbox{Crea}+ \beta_{c.8} \,\mbox{LBBB}+  \beta_{c.9}\,\mbox{PAF}+ +\beta_{c.10} \,\mbox{QRS},\nonumber \vspace*{0.5cm}\\
 &  \boldsymbol x^{\top}_{nc}\,\boldsymbol \beta_{nc}=  \,\,\beta_{nc.1}I_{Man} +   \beta_{nc.2}I_{Pace}   +\beta_{nc.3}I_{Ische} +  \beta_{nc.4}I_{III} +     \beta_{nc.5}I_{IV} +\nonumber \\
 & \hspace*{1.15cm} + \,\,\beta_{nc.6}\,\mbox{Age}+  \beta_{nc.7}\,\mbox{Crea}+ \beta_{nc.8} \,\mbox{LBBB}+  \beta_{nc.9}\,\mbox{PAF}+ +\beta_{nc.10}\,\mbox{QRS}.
\end{align*}
 

Baseline covariates Age, Crea, LBBB, PAF, and QRS describe for each individual their age,  level of creatinite, LBBB, previous AF, and QRS, respectively. The reference category for   both causes of death  is being woman, having an implanted ICD, dilated heart disease, and functional classification II. 

Under the frequentist statistics it is not usual to specify  the baseline hazard function.  This is not the case for Bayesian inference, which requires the choice of a baseline hazard function to perform a full Bayesian analysis of the problem. In this regard, we account for two different specifications of the baseline hazard function associated to each competing event: a Weibull distribution and a piecewise function. In the case of a Weibull distribution with shape $\alpha_c$ and $\alpha_{nc}$   and 
scale parameter $\lambda_c$ and $\lambda_{nc}$ for the CV and N-CV death, respectively, the baseline  hazard function for $T_C$ and $T_{NC}$ is 
\begin{align*}
 &h_{0.c}(t \mid \boldsymbol \phi_{c})   = \lambda_{c} \,\alpha_{c}\, t^{\alpha_c-1}, \nonumber\\
  &h_{0.nc}(t \mid \phi_{nc})= \lambda_{nc} \,\alpha_{nc}\, t^{\alpha_{nc}-1},\,\,\, t>0.\nonumber
  \end{align*}

\noindent This baseline hazard function is a traditional model for survival data. It is very simple and easy to interpret but lacks flexibility to represent non-monotonic risks. 

Piecewise functions are defined through polynomial functions. They provide flexible modeling that can include  monotonic and non-monotonic behaviors. It is one of the traditional alternatives to the Weibull modelling used in Bayesian inference. We use the piecewiese constant function for both survival times which we define as:
\begin{align}
 & h_{0.c}(t \mid \boldsymbol \phi_{c})=   \mbox{$\sum$}_{r=1}^{R_c}\,\gamma_{c.r} \,I_{{(a_{c.r-1}, a_{c.r} ]}},\nonumber\\
 & h_{0.nc}(t \mid \boldsymbol \phi_{nc})=   \mbox{$\sum$}_{r=1}^{R_{nc}}\,\gamma_{nc.r} \,I_{{(a_{nc.r-1}, a_{nc.r} ]}},
 \label{eqn:piecewise}
 \end{align}
\noindent where $0=a_{c.0}\leq a_{c.1}\leq \ldots \leq a_{c.R_c}$ and $0=a_{nc.0}\leq a_{nc.1}\leq \ldots \leq a_{nc.R_{nc}}$ is a partition of the relevant time axis for CV and N-CV death, respectively. In this case, the indicator function $I_A$ is 1 if $t$ belongs to interval $A$ and zero otherwise. In our study, we have considered   partitions determined by knots 0, 3, 6, 12, 24, 36, 48, and 60 month for CV death and knots 0, 24 and 48 months for N-CV death.  They have been selected by the cardiologist who collaborates in the project who is also a co-author of this paper.

\begin{table}[htb]
\begin{center}
\caption{Approximate posterior distribution for the  full model parameters, considering a Weibull and a piecewise function for the baseline hazard function.}
\begin{tabular}{lcccc}
\hline
&  \multicolumn{2}{c}{\textbf{Weibull baseline}} & \multicolumn{2}{c}{\textbf{Piecewise baseline}}\\
Covariates &  Mean & SD & Mean & SD\\
\hline
 \multicolumn{4}{l}{\textbf{Cardiovascular death }}\\
 Weibull shape  & 0.785 & 0.091 & --- & --- \\
 Weibull scale  & 0.009 & 0.006 & --- & --- \\
 \vspace{-0.3cm} \\
Age &   0.301 & 0.176 & 0.268 & 0.174 \\
Creatinine &   0.493 & 0.126 & 0.431 & 0.121 \\
CRT Device (Pacemaker)  & 0.663 & 0.311 & 0.662 & 0.304 \\
Etiology (Ischemic)   & 0.585 & 0.311 & 0.590 & 0.302 \\
LBBB &   -0.019 & 0.291 & -0.101 & 0.295 \\
NYHA (Class III)&   1.221 & 0.639 & 1.119 & 0.642 \\
NYHA (Class IV)&   1.734 & 0.755 & 1.793 & 0.750 \\
Previous AF & 0.315 & 0.290 & 0.309 & 0.290 \\
 QRS &  0.008 & 0.139 & -0.001 & 0.137 \\ 
 Sex (Man)    & 0.322 & 0.457 & 0.221 & 0.452 \\

\hline 
\multicolumn{5}{l}{\textbf{Non-Cardiovascular death }}\\
Weibull shape&     1.234 & 0.166 & --- & --- \\
Weibull scale &   0.007 & 0.005 & --- & --- \\
 \vspace{-0.3cm} \\
Age &   0.688 & 0.242 & 0.577 & 0.233 \\
Creatinine &   0.306 & 0.242 & 0.120 & 0.226 \\
CRT Device (Pacemaker)&   -0.096 & 0.364 & -0.212 & 0.360 \\
Etiology (Ischemic) &  0.387 & 0.381 & 0.415 & 0.377 \\
LBBB &   0.518 & 0.369 & 0.570 & 0.369 \\
NYHA (Class III)&   0.527 & 0.589 & 0.660 & 0.572 \\
NYHA (Class IV)&   -1.316 & 1.406 & -0.997 & 1.400 \\
Previous AF   & 0.556 & 0.367 & 0.623 & 0.357 \\
 QRS &  0.199 & 0.170 & 0.179 & 0.163 \\
 Sex (Man) &   0.092 & 0.427 & 0.047 & 0.422 \\
 \hline 
\label{tab:2}
\end{tabular}
\end{center}
\end{table}

We need to complete the Bayesian model with the specification of a prior distribution for all the parameters of the model, $\boldsymbol \theta=(\boldsymbol \phi_c, \boldsymbol \phi_{nc}, \boldsymbol \beta_c, \boldsymbol \beta_{nc})$. We assume prior independence and have selected, as much as possible, non-informative prior distributions that are standard in this application environment  in order  to give all the prominence of the inferential process to the data. We have elicited  normal distributions $N(0,0.001)$ for the coefficients of regression in $\boldsymbol \beta_c$, $\boldsymbol \beta_{nc}$. In the case of Weibull baseline hazard functions, we have elicited gamma distributions $\text{Ga}(0.01, 0.01)$ for the $\lambda$'s  and uniforms $\text{U}(0, 10)$ for the $\alpha$'s. Alternatively, in the case of the hazard functions defined through the piecewise functions in  (\ref{eqn:piecewise}), we have selected gamma distributions $\text{Ga}(0.01, 0.01)$ for the $\phi$'s coefficients. 
We are working with two models that are identical except for the specification of the baseline hazard function. The   posterior distribution 
  $\pi(\boldsymbol{\theta}\mid \mathcal{D})$ of each of them has been approximated  via Bayesian inference through MCMC with JAGS software  (Plummer, 2003) through the {\tt{rjags}} package (Plummer, 2023) in the {\tt{R}} version 4.1 (R Core Team, 2023).  Three MCMC chains, 100,000 iterations, 1,000 burning iterations and a thinning interval of 10 have been used to compute each model (Álvares \textit{et al.}, 2021). For each model, convergence has been checked via the Gelman-Rubin statistic.

Table \ref{tab:2} includes a summary of the approximate  posterior distribution of the parameters of both models (whose only formal difference is in the specification of the hazard function). In particular, the  posterior mean and standard deviation of the coefficients of both models are presented. Regarding the results, we would first like to comment on the similarity between the estimates of the parameters associated with the different covariates in both models, both for CV and N-CV death. It is also interesting to note that in both types of death, the  posterior standard deviation associated with the categorical covariates is greater than that estimated for the quantitative covariates and that greater uncertainty is observed in the Cr death group. In the case of CV death, the categories of being a man, have a pacemaker device, an ischaemic etiology of the disease, and a functional classification   III and IV generate an increase in the hazard function and, therefore,  lower survival times.  Quantitative covariates also have a negative impact on survival, except for LBBB and certainly QRS. In the N-CV death group, it is worth noting the low relevance of gender and the type of device implanted. In addition, note   that  having a diagnosis in functional classification IV decreases the risk of death, but age  increases this risk notably.

Some alternative prior distributions on model parameters have been tested in order to check the stability of the obtained posterior distribution: $N(0,0.01)$ and a $U(-10,10)$ for the regression coefficients. Regarding the coefficients of the piecewise function, two additional prior specifications have been considered as described in Lázaro \textit{et al.} (2020):

\begin{enumerate}
\item $\pi(\gamma_{c.r})=\mbox{Ga}(\omega_{c.0} \,\eta_{c.0}(a_{c.r}-a_{c.r-1}),\, \omega_{c.0}  (a_{c.r}-a_{c.r-1}))$ and $\pi(\gamma_{nc.r})=\mbox{Ga}(\omega_{nc.0} \,\eta_{nc.0}(a_{nc.r}-a_{nc.r-1}), \,\omega_{nc.0} (a_{nc.r}-a_{nc.r-1}))$, where $\omega_{c.0}=\omega_{nc.0}=0.01$, $\eta_{c.0}=0.69315/\tilde{t}_{c}$, and $\eta_{nc.0}=0.69315/\tilde{t}_{nc}$ being $\tilde{t}_c$, and $\tilde{t}_{nc}$ the median survival time for the cardiovascular and non-cardiovascular cause, 1.79 and 2.38 years, respectively.
\item $\pi(\gamma_{c.r} \mid \gamma_{c.1}, \,\gamma_{c.2}, \ldots, \gamma_{c.r-1})=\mbox{Ga}(\eta_{c.0}, \eta_{c.0}/ \gamma_{c.r-1}  )$ and $\pi(\gamma_{nc.r} \mid \gamma_{nc.1}, \,\gamma_{nc.2}, \ldots, \gamma_{nc.r-1})=\mbox{Ga}(\eta_{nc.0}, \,\eta_{nc.0}/ \gamma_{nc.r-1}  )$, where $\eta_{c.0}=\eta_{nc.0}=0.01$ and 
  $\pi(\gamma_{c.1})=\pi(\gamma_{nc.1})=\text{Ga}(0.01,0.01)$.  \end{enumerate}

We have not detected any relevant change in the posterior distribution obtained from these new prior distributions compared with the results presented in Table \ref{tab:2}. Furthermore, fitting metrics (DIC, WAIC and LPML) have been obtained for all the models considered, yielding similar results. In the light of the robustness of the posterior distribution obtained, we decided to keep the simpler and more "standard" prior distribution for the regression coefficients: a $N(0,0.001)$ prior. Regarding the specification for the baseline piecewise function, and as it can be checked in the following sections, choosing a Weibull distribution instead always improves the model fit.


  \subsection{Model uncertainty}

   \begin{figure}[htb]
\begin{center}
\includegraphics[scale=0.35]{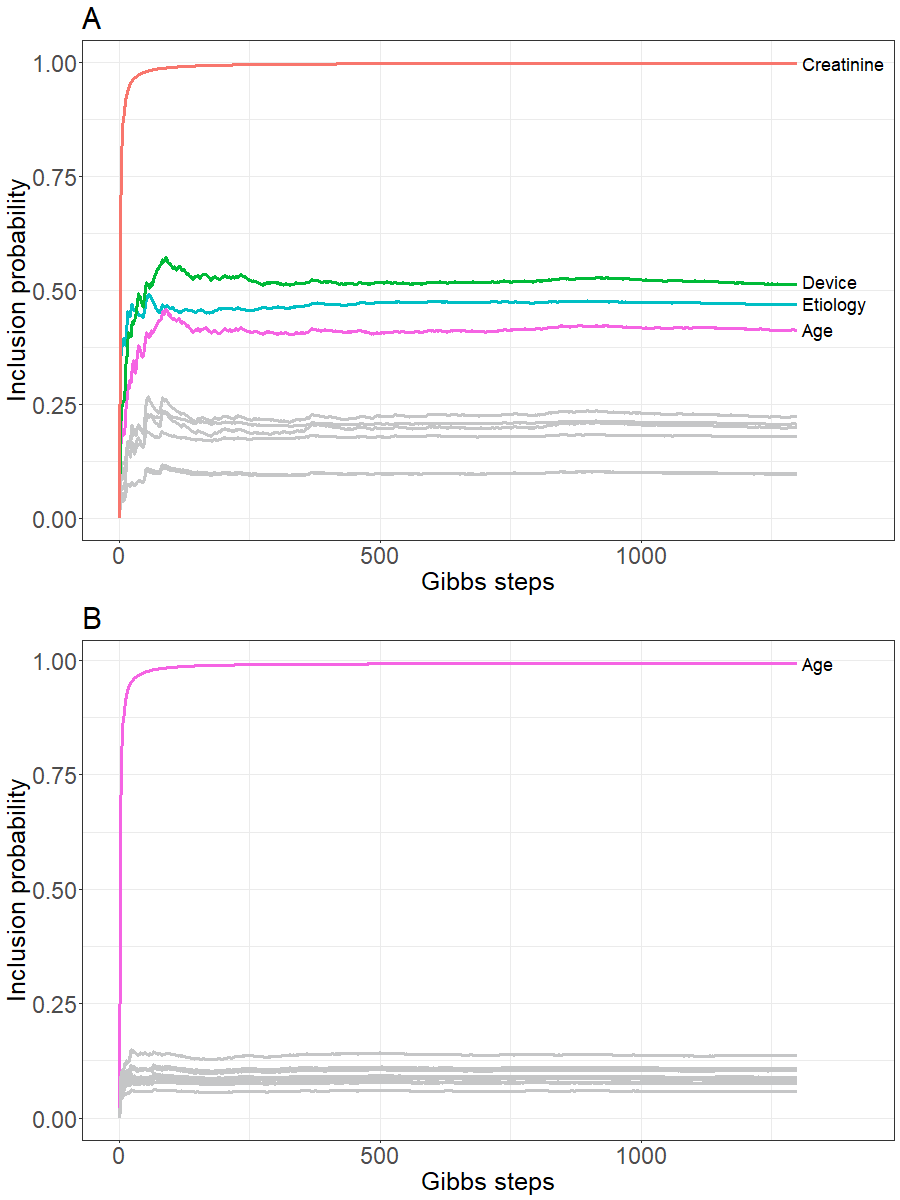}
\caption{Estimated inclusion probabilities of each of the covariates in the model using the Bayes factors based procedure in relation to the number of Gibbs iterations. Both competing events, (A) cardiovascular and (B) non-cardiovascular death, are considered.}
\end{center}
\label{figure:2}
\end{figure}

As we have introduced in Section 3, we have used two variable selection procedures: the proposal of Garcia-Donato et al. (2023) through the Bayes factor and the spike-and-slab method. Regarding the first proposal, it should be remembered that it does not consider the baseline hazard function and only takes into account the covariates. A total number of 1000 Gibbs iterations and a burn-in period of 300 were used. The full model contained 10 covariates and the  initial model dimension was set as 5. The stability of the inclusion probabilities was approximately achieved on the first 300 iterations as seen in Figure 3. The high probability of inclusion of Creatinine and Age in the case of CV and N-CV, respectively, can be observed. The probability of inclusion of the remaining covariates is very low for N-CV death. When it comes to CV death we can observe inclusion probabilities around 0.5 for CRT Device, Etiology and Age and very small for the remaining covariates.   

Table \ref{tab:3} shows the five competing risks models that the García-Donato et al. (2023)  procedure selects with the highest probability. We again confirm that the results obtained for both causes of death are very different. We have little uncertainty about the best model in the case of
N-CV death, with only covariate Age; and a little more uncertainty in the case of CV death: Creatinine appears in all the models and, in fact, the model with the highest probability only includes this covariate. CRT Device and Etiology appear in three of the five models and Age in only one. 

    \begin{table}[htb]
\begin{center}
\caption{Posterior model probabilities of the five  most probable competing risks models}
\begin{tabular}{rl}
\hline
Model & Post. probability\\
\hline
\multicolumn{1}{r}{\textbf{Cardiovascular death}}\\
Creatinine & 0.1913\\
Creatinine + Etiology & 0.1452\\
Creatinine + CRT Device + Age + Etiology & 0.1113\\
Creatinine + CRT Device & 0.0815\\
Creatinine + CRT Device + Etiology & 0.0465\\
\hline
\multicolumn{1}{r}{\textbf{Non-cardiovascular death}}\\
Age & 0.6075\\
Age + NYHA(IV) & 0.0790\\
Age + NYHA(III) & 0.0592\\
Age + Etiology & 0.0493\\
Age + Sex & 0.0423\\
\hline
\label{tab:3}
\end{tabular}
\end{center}
\end{table}

 The spike and slab method applies to all elements in the model, the baseline hazard function and the covariates. We recall that we have worked with two different formulations of the baseline hazard function, one through the Weibull distribution and the second through a piecewise-defined function.   A $N(0,0.001)$ prior distribution  has been selected for modeling each slab term and the beta distribution    Beta(1,1) has been elicited associated to probability  $\eta$ in (\ref{eqn:spike}).

\begin{table}[htb]
\begin{center}
\caption{Inclusion probabilities   of covariates associated with cardiovascular and non-cardiovascular death determined through the Bayes factor and the spike and slab method, this latter with a Weibull and a piecewise baseline hazard function. }
\begin{tabular}{lcccccc}
\hline
   & & & \multicolumn{4}{c}{\textbf{Spike-and-slab prior}}\\
& \multicolumn{2}{c}{\textbf{Bayes Factor}} & \multicolumn{2}{c}{Weibull} & \multicolumn{2}{c}{Piecewise}\\
& CV death & N-CV death & CV death & N-CV death& CV death & N-CV death\\
\hline
Age &   0.412 &  0.947 & 0.021 &  0.984 & 0.037 &  0.990 \\ 
Creatinine &  0.963 & 0.087 &  0.998 & 0.011 & 0.993 & 0.017 \\  
CRT Device &  0.507 & 0.059 & 0.095 & 0.010 & 0.121 & 0.008 \\
Etiology &  0.447 & 0.091 & 0.190 & 0.020 & 0.162 & 0.015 \\
LBBB & 0.099 & 0.082 & 0.012 & 0.014  & 0.009 & 0.007 \\
NYHA (Class III) & 0.225 & 0.102 & 0.033 & 0.022 & 0.055 & 0.071 \\
NYHA (Class IV) & 0.214 & 0.133 & 0.028 & 0.051 & 0.031 & 0.082 \\
Previous AF & 0.176 & 0.109 & 0.028 & 0.028 & 0.021 & 0.053 \\
QRS & 0.100 & 0.086 & 0.004 & 0.010 & 0.010 & 0.003 \\
Sex & 0.216 & 0.077 & 0.048 & 0.012 & 0.055 & 0.022 \\
\hline
\label{tab:4}
\end{tabular}
\end{center}
\end{table}

Table \ref{tab:4}   shows the estimated probability of inclusion in the model of each of the covariates provided by the Bayes factor method and by a spike distribution method, the latter with the Weibull and the piecewise baseline hazard function.  We can observe that the results obtained with the spike are much more drastic than with the procedure based on the Bayes factor because they only give a high inclusion probability estimate to the covariate Creatinine in the case of CV death and to Age in the case of N-CV death. Both covariates are also the favorites of the Bayes factor. No relevant differences were observed between the two modeling approaches based on the spike-and-slab method.  The model with the highest posterior probability and the model that includes all covariates with inclusion probability 0.5 or higher (Median posterior probability model, Barbieri and Berger, 2004) coincide in this case: Creatinine for CV death and Age for N-CV death. 
 

All the information obtained on previous analyses matches with clinical knowledge. According to the clinicans criteria, the best model would be considering Age, Creatinine, CRT Device, and Etiology, and adding the covariate NYHA functional class for CV death  due to their clinic  relevance and leaving the proposal of both methods in the case of N-CV death with only Age.  

As a final step, we used the DIC, WAIC and LPML criteria to compare four models: the full model with all covariates, the model that only includes covariates that have a probability of inclusion greater than 0.5, the model with covariates with inclusion probabilities greater than 0.4, and the previous model with the NYHA covariate added.  In all models, we have considered a Weibull and a piecewise constant baseline hazard function.  All the models have a fairly similar behavior with the three criteria, although the models with Weibull baseline hazard function are always better and also, in all cases, the last one is the best. Finally, we decided to select the latter model: Weibull baseline hazard function for all events and   covariates. Age, Creatinine, Device, Etiology and NYHA for CV death, and   Age for N-CV death. This one is the model recommended by the cardiologist and the one with slightly better fitting metrics, as it can be seen in Table \ref{tab:5}.

\begin{table}[htb]
\begin{center}
\caption{Fitting metrics (DIC, WAIC and LPML) for all the models considered. M1 refers to creatinine for CV death and age for NCV death. M2 refers to creatinine, etiology, DAI and age for CV death and age for NCV. M3 refers to creatinine, etiology, DAI, NYHA and age for CV death and age for NCV.}
\begin{tabular}{llll}
\hline
& \textbf{DIC} & \textbf{WAIC} & \textbf{LPML}\\
\hline
 \textbf{Weibull baseline.} & & &\\
 Complete  & 59200698 & 59200726 & -367.97 \\
 M1  & 59200711 & 59200718 & -359.25 \\
 M2  & 59200700 & 59200710 & -355.44 \\
 M3  & 59200696 & 59200709 & -355.25 \\
 \vspace{-0.25cm} \\
 \textbf{Piecewise constant baseline.} & & & \\
 Complete  & 59200917 & 59200934 & -469.61 \\
 M1  & 59200931 & 59200937 & -470.94 \\
 M2  & 59200920 & 59200929 & -465.56 \\
 M3  & 59200915 & 59200925 & -464.06 \\
\hline 
\label{tab:5}
\end{tabular}
\end{center}
\end{table}

\subsection{Selected model and posterior outputs}

Table \ref{tab:6} shows a summary of the approximated posterior distribution for the parameters of the selected model. We have used JAGS in the same way as with the full model to obtain an approximate sample from the subsequent posterior distribution. If we observe the results of the complete model and the selected model, we can see stability and similarity in the information corresponding to the common covariates in both events, especially for CV death. In this case, the ischemic etiology is associated with an increased risk of death compared to the dilated etiology. This also occurs with the CRT pacemaker with regard to the defibrillator device. Age, Creatinine and NYHA functional classification III and mainly IV have also a positive impact on CV death. Age is the only relevant covariate for N-CV death, with a strong positive impact on the risk of death.

\begin{table}[htb]
\begin{center}
\caption{Summary of the approximate posterior distribution for the selected model.}
\begin{tabular}{lllll}
\hline
& Mean & SD & Median & CI 95\%\\
\hline
\multicolumn{5}{l}{\textbf{Cardiovascular death}}\\
Weibull shape & 0.780 & 0.090 & 0.777 & [0.611, 0.965] \\
Weibull scale& 0.012 & 0.007 & 0.010 & [0.002, 0.030] \\
 \vspace{-0.3cm} \\
  Age  & 0.326 & 0.172 & 0.323 & [-0.004, 0.674] \\
  Creatinine   & 0.505 & 0.120 & 0.505 & [0.267, 0.736] \\
  CRT Device  (Pacemaker) & 0.689 & 0.316 & 0.687 & [0.087, 1.315] \\
  Etiology (Ischemic) & 0.616 & 0.287 & 0.616 & [0.055, 1.187] \\
 NYHA (Class III)  & 1.284 & 0.644 & 1.232 & [0.177, 2.712] \\
 NYHA (Class IV)  & 1.753 & 0.769 & 1.718 & [0.345, 3.375] \\
\hline
\multicolumn{5}{l}{\textbf{Non-Cardiovascular death}}\\
Weibull shape & 1.218 & 0.158 & 1.214 & [0.920, 1.537] \\
Weibull scale & 0.022 & 0.007 & 0.021 & [0.011, 0.039] \\
 \vspace{-0.3cm} \\

Age & 0.826 & 0.224 & 0.819 & [0.403, 1.283] \\
\hline
\label{tab:6}
\end{tabular}
\end{center}
\end{table}


\begin{figure}[htb]
\begin{center}
\includegraphics[scale=0.42]{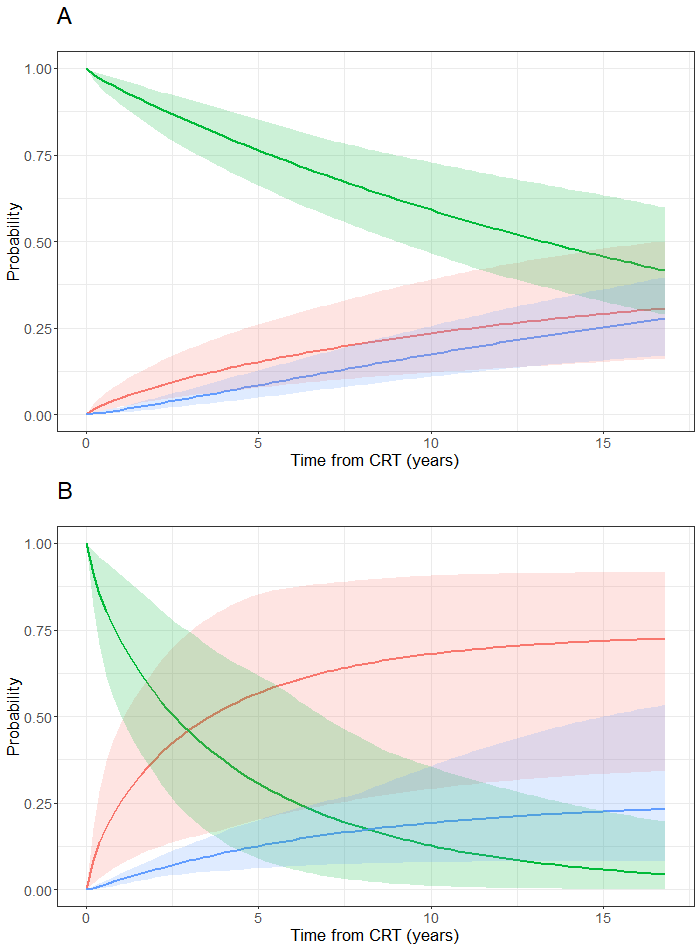}
\caption{Posterior mean and 95\% credible intervals of transition probabilities to CV death (red), to N-CV death (blue) and survival (green) for two different patient profiles. (A) 65 year-old patient with serum creatinine of 1.2 mg/dL, ischemic etiology, defibrillator implanted and NYHA III; (B) 75 year-old patient with serum creatinine of 1.5 mg/dL, ischemic etiology, pacemaker implanted and NYHA IV.}
\end{center}
\label{figure:3}
\end{figure}

Figure 4 shows the posterior mean for the relevant  transition probabilities of the model: from the initial state defined by the time in which the CRT implant is performed until CV, N-CV death or the permanence in that initial state in which the patient continues with life. We have computed these probabilities for two different patient profiles. On the one hand, 65 year-old patients, with ischemic disease etiology, pacemaker implanted, NYHA class III and a low baseline creatinine level (1.2 mg/dL ). In the second group of patients we included 75 year-old patients, ischemic etiology, pacemaker, NYHA class IV and a high creatinine level (1.5 mg/dL). From now on, we will call the patients in the first group type A patients and the second group type B patients. 

The behavior of transition probabilities in both types of patient profiles is very different. Patients with a type A profile have better prognoses than those with a B profile.   The probability of staying alive after CRT implantation in patients with profile A is greater than 0.5 until approximately 13 years after implantation, whereas for patients with profile B it is only greater than 0.5 until approximately 3 years after implantation.   The probability of being alive 5, 10  and 15 years after implantation is around 0.76, 0.57, and 0.46 in type A patients, and around 0.31, 0.13, and 0.05 in type B patients, respectively. These probabilities are also known   in clinical terms as 5, 10 and 15-year survival. The probability of CV death before 5, 10 or 15 years after implantation is around 0.15, 0.25 and 0.29 in type A patients. These probabilities become 0.57, 0.67 and 0.73 in type B patients. As for N-CV death, in type A patients there is an estimated probability of 0.09, 0.18 and 0.23 that death will occur in the first 5, 10 and 15 years after the implant. These estimates are 0.12, 0.21 and 0.22 in type B patients.  There do not seem to be many differences in the evolution of N-CV death in both patient profiles. The largest differences are found in the survival and CV death probabilities.

Figure 5 includes, for patients type A and B, the posterior mean of the conditional transition probabilities to a CV or a N-CV death as well as the permanence as a living individual in a certain period of time given  that he/she has been implanted and has survived a known length of time after implantation. The term conditional here is justified by the assumption that the patient continues to live for a certain period of time  after implantation, in our case 3, 5 and 8 years. If we look at type A patients, we recall that their 10 and 15-year survival was around 0.57 and 0.46, respectively. If after 3, 5 or 8 years a type A patient had survived, the expected posterior probability that he would still be alive 10  years after implantation increases to  around 0.70, 0.76 and 0.90, respectively.  These expected probabilities for a 15-years period of time are about 0.55, 0.60 and 0.73, respectively. 
\begin{figure}[htb]
\begin{center}
\includegraphics[scale=0.40]{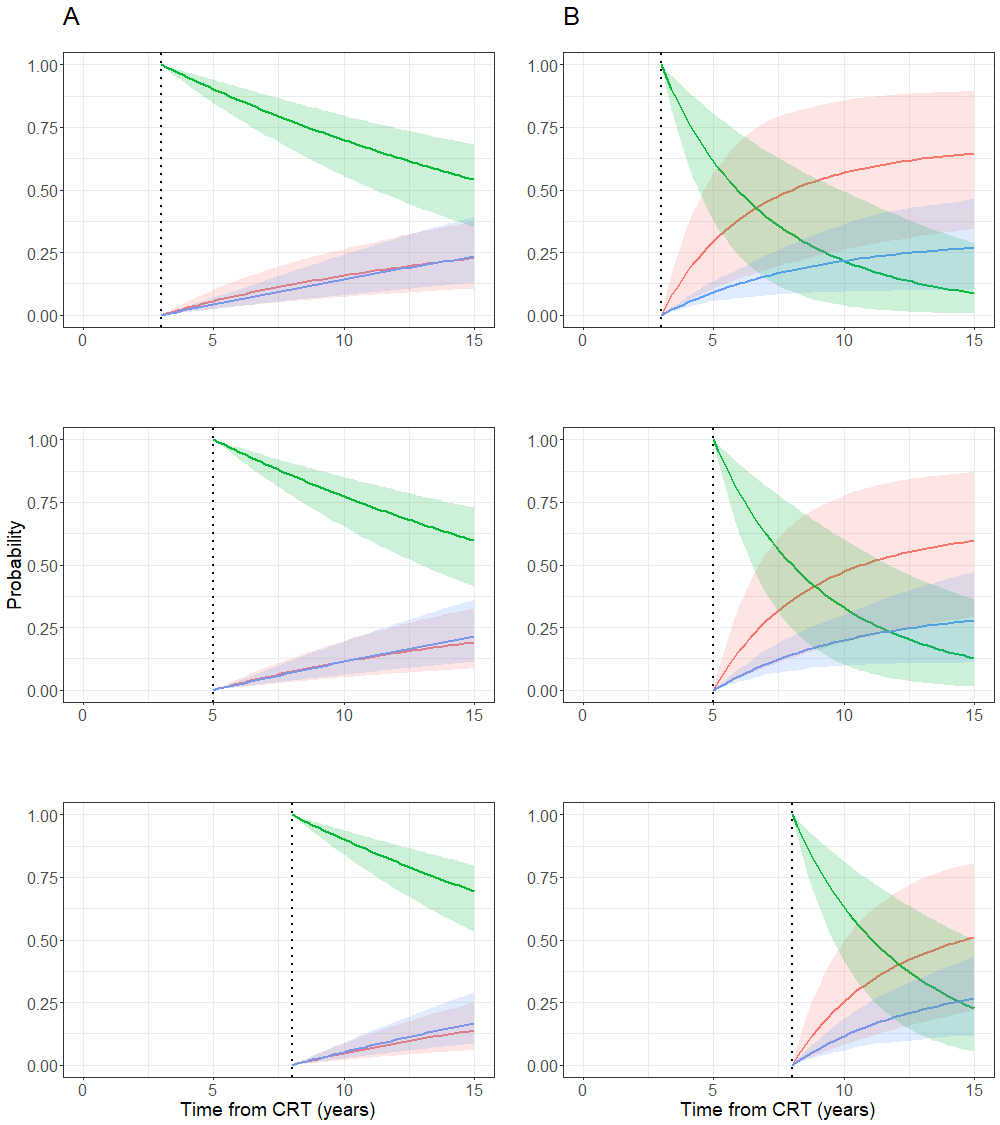}
\caption{Posterior mean and 95\% credible intervals of conditional transition probabilities to cardiovascular death (red line), to non-cardiovascular death (blue line) and survival (green line) for two different patient profiles. 3, 5 and 8 years are considered as the initial survival time. (A) 65 year-old patient with serum creatinine of 1.2 mg/dL, ischemic etiology, defibrillator implanted and NYHA III; (B) 75 year-old patient with serum creatinine of 1.5 mg/dL, ischemic etiology, pacemaker implanted and NYHA IV. }
\label{figure:4}
\end{center}
\end{figure}
In the case of CV death the survival of type A patients was less than 10 and 15 years with probabilities 0.24 and 0.29. These probabilities turn out to be 0.67 and 0.73 for type B patients. Adding the condition that the patient is still alive 3 years after implantation, these probabilities become 0.15 and 0.23 for type A patients and 0.54 and 0.62 for type B patients. respectively. In the case of having lived more than 5 years after transplantation, the probabilities of living less than 10 years become 0.12 and 0.17 for type A patients, and 0.46 and 0.61 for type B patients. Finally, for patients who survive more than 8 years after implantation, the probability of a CV-death less than 10 or 15 years after implantation is 0.05 and 0.13 for type A patients and 0.24 and 0.50 for type B patients, respectively.  

The conditional probabilities associated with N-CV death are the most stable.  We know that in type A patients the probability of N-CV death before 10 to 15 years after implantation is around 0.19 and 0.25, and approximately 0.21 and 0.23 in type B patients. If we add the condition of having survived more than 3, 5 and 8 years after implantation these probabilities become 0.15 and 0.23, 0.12 and 0.23, and    0.05 and 0.14 in type A patients, respectively. In patients type B, the aforementioned probabilities are approximately  0.23 and 0.26, 0.21 and 0.26, and   0.11 and 0.25, respectively.  

\section{Conclusions}

We have proposed a Bayesian approach to competing risks survival models through
the so-called cause-specific hazard models which we have complemented through a multistate model perspective. We have carried out an inferential process with non-informative prior distributions and MCMC methods to approximate the respective posterior distribution.   Outputs of interest in the statistical study also  include the posterior distribution for the overall survival function of the time in which the first event occurs and  the cumulative incidence function of that survival time due to each event-cause. In addition, we have  discussed the posterior distribution for the transition probabilities between the different states of the multistate model both from the initial event  and the case where the individual under consideration remains under the initial situation for a certain period of time. 

 Model uncertainty  has been  adressed for  both variable selection and the specification of the baseline hazard function  in the subsequents proportional hazard functions associated to the time in which the event occurs for each cause.  The proposal by 
Garc\'ia-Donato et al (2023) with regard Bayes factor for univariate Cox regression  is generalised to competing risks models. An extension of the CPO, which is the basis for logarithm of the pseudomarginal likelihood criterion based on the predictive power of the model,   for competing risks scenarios is also presented.

We have applied all the presented procedures to data from a retrospective follow-up study conducted in a cardiac institution in Galicia (Spain) between January 2005 and April 2015 to assess time to CV and N-CV death in patients undergoing CRT. The results obtained have been very positively assessed by the cardiologist of the team insofar as they could be very useful in clinical practice. It is worth highlighting  the selection of the set of covariates chosen in the survival times associated with both causes of death and the information provided by the posterior distribution related to the relevant transition probabilities for different patient profiles. 

\section*{Acknowledgement}

This research was supported by the Xunta de Galicia via an industrial doctoral grant to the first author (Doutoramento Industrial 2021, IN606D), and  the Ministerio de Ciencia, Innovación y Universidades through the project \\“PID2022-136455NB-I00 MCIN/AEI/10.13039/501100011033/FEDER, UE" and the European Regional Development Fund.

\section*{Bibliography}

[1] Alvares, D., Lazaro, E., Gomez-Rubio, V., and Armero C. (2021). Bayesian survival analysis with BUGS. Statis-
tics in Medicine 40(12), 2975-3000.

[2] Alvarez-Alvarez, B., Garcıa-Seara, J., Martınez-Sande, J. L., Rodrıguez-Mañero, M., Fernandez Lopez, X. A.,
Gonzalez-Melchor, L., Iglesias-Alvarez, D., Gude, F., Dıaz-Louzao, C., \& Gonzalez-Juanatey, J. R. (2021).
Long-term cardiac reverse remodeling after cardiac resynchronization therapy. Journal of Arrhythmia 37(3),
653–659.

[3] Ameis, L., Kuss, O., Hoyer, A., \& Mollenhoff, K. (2024). A non-parametric proportional risk model to assess a
treatment effect in time-to-event data. Biometrical Journal, 66(4). https://doi.org/10.1002/bimj.202300147

[4] Andersen, P. K. and Keiding N. (2002). Multi-state models for event history analysis. Statistical Methods in
Medical Research 11(2), 91-115.

[5] Barbieri, M. M. and Berger, J. O. (2004). Optimal predictive model selection. The Annals of Statistics 32(3),
870–897.

[6] Barbieri, M. M., Berger, J. O., George, E. I., and Rockova, V. (2021). The Median Probability Model and
Correlated Variables. Bayesian Analysis 16(4), 1085–1112.

[7] Basu, S., \& Tiwari, R. C. (2009). Breast Cancer Survival, Competing Risks and mixture Cure Model: A
Bayesian analysis. Journal of the Royal Statistical Society Series a (Statistics in Society), 173(2), 307–329.
https://doi.org/10.1111/j.1467-985x.2009.00618.x

[8] Borchers, H. (2022). pracma: Practical Numerical Math Functions. R package version 2.4.2, https://CRAN.R-
project.org/package=pracma

[9] Bristow, M. R., Saxon, L. A., Boehmer, J., Krueger, S., Kass, D. A., De Marco, T., Carson, P., DiCarlo, L.,
DeMets, D., White, B. G., DeVries, D. W., \& Feldman, A. M. (2004). Cardiac-Resynchronization Therapy with
or without an Implantable Defibrillator in Advanced Chronic Heart Failure. New England Journal of Medicine
350(21), 2140–2150.

[10] Chen, M.-H., de Castro, M., Ge, M., and Zhang, Y. (2014). Bayesian Regression Models for Competing Risks.
In: Klein JK, van Houwelingen HC, Ibrahim JG, and Scheike TH (eds) Handbook of Survival Analysis. Boca
Raton, FL: Chapman \& Hall/CRC, 179-198.

[11] Cleland, J. G., Abraham, W. T., Linde, C., Gold, M. R., Young, J. B., Claude Daubert, J., Sherfesee, L., Wells,
G. A., \& Tang, A. S. L. (2013). An individual patient meta-analysis of five randomized trials assessing the ef-
fects of cardiac resynchronization therapy on morbidity and mortality in patients with symptomatic heart failure.
European Heart Journal 34(46), 3547–3556.

[12] Cleland, J. G. F., Daubert, J.-C., Erdmann, E., Freemantle, N., Gras, D., Kappenberger, L., \& Tavazzi, L. (2005).
The Effect of Cardiac Resynchronization on Morbidity and Mortality in Heart Failure. New England Journal of
Medicine 352(15), 1539–1549.

[13] Gail M. (1975). A review and critique of some models used in competing risk analyses. Biometrics 31, 1726-
1736.

[14] Garcıa-Donato, G., Cabras, S., and Castellanos, M. E. (2023). Model uncertainty quantification in Cox regression.
Biometrics 79(3), 1726-1736.

[15] Gaynor JJ, Feuer EJ, and Tan CC. (1993). On the use of cause-specific failure and conditional failure probabilities:
examples from clinical oncology data. Journal of the American Statistical Association 88, 400–409.

[16] Ghio, S., Freemantle, N., Scelsi, L., Serio, A., Magrini, G., Pasotti, M., Shankar, A., Cleland, J. G. F., \& Tavazzi,
L. (2009). Long-term left ventricular reverse remodelling with cardiac resynchronization therapy: results from
the CARE-HF trial. European Journal of Heart Failure 11(5), 480–488.

[17] Gray, R. J. (1988). A class of K-Sample tests for comparing the cumulative incidence of a competing risk. The
Annals of Statistics, 16(3). https://doi.org/10.1214/aos/1176350951

[18] Gray, B. (2022). cmprsk: Subdistribution Analysis of Competing Risks. R package version 2.2-11,
https://CRAN.R-project.org/package=cmprsk.

[19] Ibrahim, J.G., Chen, MH. \& Sinha, D. (2001). Model Comparison. In Bayesian Survival Analysis, 208-261.
Springer Series in Statistics. Springer, New York.

[20] Lazaro, E., Armero, C., and Alvares, D. (2020). Bayesian regularization for flexible baseline hazard functions in
Cox survival models. Biometrical Journal 63(1), 7-26.

[21] Laud, P. W. (2014). Bayesian Model Selection. In Klein, J., van Houwelingen, H., I. J. S. T. (Eds.), Handbook of
survival analysis (pp. 285-299). Boca Raton, FL: Chapman \& Hall.

[22] Marijon, E., Leclercq, C., Narayanan, K., Boveda, S., Klug, D., Lacaze-Gadonneix, J., Defaye, P., Jacob, S.,
Piot, O., Deharo, J.-C., Perier, M.-C., Mulak, G., Hermida, J.-S., Milliez, P., Gras, D., Cesari, O., Hidden-Lucet,
F., Anselme, F., Chevalier, P., Le Heuzey, J.-Y. (2015). Causes-of-death analysis of patients with cardiac resyn-
chronization therapy: an analysis of the CeRtiTuDe cohort study. European Heart Journal 36(41), 2767–2776.

[23] Mitchell, T. and Beauchamp, J. (1988). Bayesian variable selection in linear regression. Journal of the American
Statistical Association, 83(404), 1023–32.

[24] Moss, A. J., Hall, W. J., Cannom, D. S., Klein, H., Brown, M. W., Daubert, J. P., Estes, N. A. M., III, Foster,
E., Greenberg, H., Higgins, S. L., Pfeffer, M. A., Solomon, S. D., Wilber, D., \& Zareba, W. (2009). Cardiac-
Resynchronization Therapy for the Prevention of Heart-Failure Events. New England Journal of Medicine
361(14), 1329–1338.

[25] Ntzoufras, I. (2008). The Predictive Distribution and Model Checking. In Bayesian Modeling Using WinBUGS,
pp. 341-388. https://doi.org/10.1002/9780470434567.ch10

[26] Pepe, Margaret Sullivan, and Motomi Mori. “Kaplan—Meier, Marginal or Conditional Probability Curves in
Summarizing Competing Risks Failure Time Data?” Statistics in medicine. 12.8 (1993): 737–751. Web.

[27] Plummer, M. (2003). JAGS: A Program for Analysis of Bayesian Graphical Models Using Gibbs Sampling.
Proceedings of the 3rd International Workshop on Distributed Statistical Computing (DSC 2003), Vienna, 20-22
March 2003, 1-10.

[28] Plummer M. (2023). rjags: Bayesian Graphical Models using MCMC. R package version 4-14, ¡https://CRAN.R-
project.org/package=rjags¿.

[29] R Core Team (2023). R: A language and environment for statistical computing. R Foundation for Statistical
Computing, Vienna, Austria. https://www.R-project.org/.

[30] Spiegelhalter, D., Best, N., Carlin, B. P., \& Van Der Linde, A. (2002). Bayesian measures of model complexity
and fit. Journal of the Royal Statistical Society Series B: Statistical Methodology 64(4), 583-639

[31] Watanabe, S. (2010). Asymptotic Equivalence of Bayes Cross Validation and Widely Applicable Information
Criterion in Singular Learning Theory. Journal of Machine Learning Research 11, 3571–3594.

[32] Yu, J., \& Huang, Y. (2023). Unified semicompeting risks analysis of hepatitis natural history through mediation
modeling. Statistics in Medicine, 42(24), 4301–4318. \\https://doi.org/10.1002/sim.9862

\end{document}